\def\be{\begin{equation}}
\def\ee{\end{equation}}
\def\bea{\begin{eqnarray}}
\def\eea{\end{eqnarray}}
\def\bse{\begin{subequations}}
\def\ese{\end{subequations}}

\documentclass[prl,twocolumn,showpacs,amsmath,amssymb]{revtex4}
\usepackage{graphicx}
\usepackage{dcolumn}
\usepackage{bm}
\begin{document}
\title{Tricritical behavior in itinerant quantum ferromagnets
}
\author{D. Belitz$^1$, T.R. Kirkpatrick$^2$, and J{\"o}rg Rollb{\"u}hler$^1$}
\affiliation{$^1$Department of Physics and Materials Science Institute,
                  University of Oregon, Eugene, OR 97403\\
             $^2$Institute for Physical Science and Technology, and Department of
                 Physics\\
                 University of Maryland, College Park, MD 20742
         }
\date{\today}

\begin{abstract}
It is shown that the peculiar features observed in the low-temperature phase
diagrams of ZrZn$_2$, UGe$_2$, and MnSi can be understood in terms of a simple
mean-field theory. The nature of the ferromagnetic transition changes from
second order to first order at a tricritical point, and in a small external
magnetic field surfaces of first-order transitions emerge which terminate in
quantum critical points. This field dependence of the phase diagram follows
directly from the existence of the tricritical point. The quantum critical
behavior in a nonzero field is calculated exactly.
%
\end{abstract}


\maketitle

\begin{figure}[t]
\includegraphics[width=7.0cm]{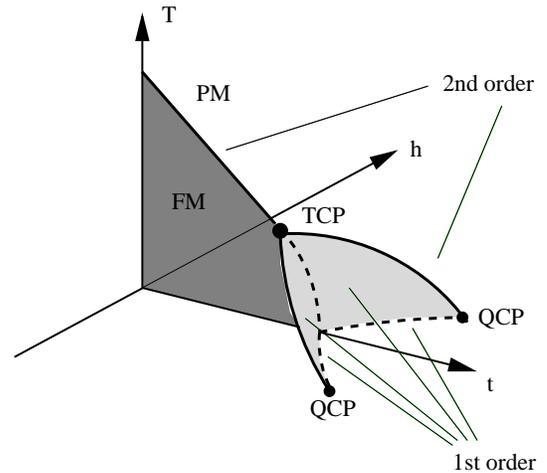}
\caption{\label{fig:1} Schematic phase diagram in the
temperature-pressure-magnetic field ($T-p-h$) space. Shown are the
ferromagnetic (FM, dark shaded) and paramagnetic (PM) phases at $h=0$, the
tricritical point (TCP), and the two quantum critical points (QCP). Also shown
are various lines of first (dashed lines) and second-order (solid lines) phase
transitions, and the ``wing'' surface of first order transitions (light
shaded).}
\end{figure}

The ferromagnetic transition at the Curie points of the elements Fe, Ni, and Co
is one of the best-known examples of a second-order phase transition. It is
well understood in terms of the band theory of metals in conjunction with the
theory of phase transitions \cite{Jones_March_1973}. Recent experimental
studies of ferromagnetic compounds with much lower Curie temperatures, among
them ZrZn$_2$ \cite{Uhlarz_et_al_2004}, UGe$_2$
\cite{Huxley_Sheikin_Braithwaite_2000}, and MnSi
\cite{Pfleiderer_Julian_Lonzarich_2001, heli_footnote}, show enigmatic behavior
which does not seem to fit into this well-established picture: If the low Curie
temperature is further decreased by means of pressure-tuning, the nature of the
transition changes from second order to first order at a tricritical point, and
in a small external magnetic field surfaces or ``wings'' of first-order
transitions emerge which extend from the coexistence line at zero field and
terminate in quantum critical points. These regions of first-order transitions
end in lines of critical points which are reminiscent of conventional
liquid-gas critical points, and which connect the tricritical point with a
quantum critical points in the zero-temperature plane. These observations are
summarized in the schematic phase diagram shown in Fig.\ \ref{fig:1}.

This structure of the phase diagram is very remarkable, for two reasons. First,
the ferromagnetic transition in zero field at high-temperature Curie points,
most notably in the elemental ferromagnets, is invariably of second order.
Also, Hertz's theory of the quantum ferromagnetic transition at $T=0$
\cite{Hertz_1976}, and its extension to nonzero temperature \cite{Millis_1993},
predict the ferromagnetic transition to be generically of second order. Second,
the persistence of the first-order transition away from the zero-field axis,
and the existence of quantum critical points at $h\neq 0$, came as a surprise
\cite{Kimura_et_al_2004}. Yet the observed structure of the phase diagram as
sketched in Fig.\ \ref{fig:1} seems to be generic, as demonstrated by the case
of ZrZn$_2$ as the latest example, where a tricritical point emerged once
sufficiently clean samples were produced \cite{Uhlarz_et_al_2004}.

In this Letter we show that all of these observations can be explained by a
theory that takes into account the fact that, in metallic systems at low
temperatures, the particle-hole excitations characteristic of systems with a
Fermi surface couple to the fluctuations of the magnetic order parameter and
substantially change the nature of the phase transition compared to the
conventional theory \cite{Belitz_et_al_2001a, Kirkpatrick_Belitz_2003}.
Furthermore, we identify the universality classes for all finite-temperature
critical points in the phase diagram, and we determine the exact critical
behavior at the quantum critical points.

Within the framework of this theory, a mean-field theory for three-dimensional
systems is defined by a free energy density
\be
f = -h\phi + t\phi^2 + v\phi^4\ln\left(\phi^2/m_0^2 + T^2/T_0^2\right) +
u\phi^4
\label{eq:1}
\ee
in terms of an order parameter $\phi$. Here $t$ and $u,v>0$ are coefficients
analogous to those in Landau theory \cite{Landau_Lifshitz_V_1980}. For later
reference we note that within Stoner theory \cite{Stoner_1938,
Sandeman_Lonzarich_Schofield_2003}, which is a particular realization of Landau
theory, $t = 1 - \Gamma_{\text{t}}N_{\text{F}}$, with $N_{\text{F}}$ the
density of states at the Fermi surface and $\Gamma_{\text{t}}$ a microscopic
spin-triplet interaction amplitude, $u$ is proportional to the second
derivative of the density of states, and $v=0$. $m_0$ is a microscopic
magnetization (e.g., one Bohr magneton $\mu_{\text{B}}$ per volume of a unit
cell), and $T_0$ is a microscopic temperature (e.g., the Fermi temperature).
The physical value of $\phi$, which we will denote by $\varphi$, is the one
that minimizes $f$. For $h=0$, $\varphi$ is equal to the magnetization $m$, and
for this case the free energy given by Eq.\ (\ref{eq:1}) was first considered
in Ref.\ \onlinecite{Belitz_Kirkpatrick_Vojta_1999}. For $h>0$, a derivation
along the same lines as in Refs. \onlinecite{Belitz_et_al_2001a,
Kirkpatrick_Belitz_2003} shows that the magnetization is now related to the
physical value of the order parameter by
\be
m = \varphi - (4\mu_{\text{B}}^2/\Gamma_{\text{t}})h.
\label{eq:2}
\ee
The equation of state, which relates $m$, $T$, and $h$, is obtained by
minimizing $f$, which leads to
\bea
h &=& 2t\,\varphi + 4v\,\varphi^3\ln\left(\varphi^2/m_0^2 + T^2/T_0^2\right)
\nonumber\\
    &&\hskip 50pt + v\,\varphi^3\,\frac{\varphi^2/m_0^2}{\varphi^2/m_0^2 + T^2/T_0^2}
    + 4u\,\varphi^3
\label{eq:3}
\eea
in conjunction with Eq. (\ref{eq:2}).

Notice that, at $T=0$, both the free energy and the equation of state are
nonanalytic functions of the order parameter by virtue of the logarithmic term.
This is in sharp contrast to ordinary Landau theory
\cite{Landau_Lifshitz_V_1980}, where $f$ is an analytic function of $\phi$, and
it reflects the fact that the particle-hole excitations have been integrated
out to obtain a free energy in terms of the order parameter only. This is a
particular example of a more general phenomenon, see Ref.\
\onlinecite{Belitz_Kirkpatrick_Vojta_2004}.

In order to discuss the mean-field theory given by Eq.\ (\ref{eq:3}), we first
recall the solution at $h=0$ \cite{Belitz_Kirkpatrick_Vojta_1999}. There is a
tricritical point at $(t=0,T=T_{\text{tc}})$, with $T_{\text{tc}} = T_0\,
e^{-u/2v}$. At $T=0$, the transition occurs at $t=t_1=m_0^2\,e^{-1}v\,e^{-u/v}$
and is of first order with the magnetization changing discontinuously from
$m=0$ to $m=m_1=m_0\,e^{-1/2}e^{-u/2v}$. The line of first-order transitions at
temperatures $T<T_{\text{tc}}$ is determined by $f(\varphi) = f'(\varphi) = 0$
and can be given explicitly in parametric form $(0\leq s\leq 1)$:
\bse
\label{eqs:4}
\bea
t(s) &=& t_1\,(1-s)^2\, e^s,
\label{eq:4a}\\
T(s) &=& T_{\text{tc}}\, s^{1/2}\, e^{-(1-s)/2}.
\label{eq:4b}
\eea
\ese

\begin{figure}[t]
\includegraphics[width=6.0cm]{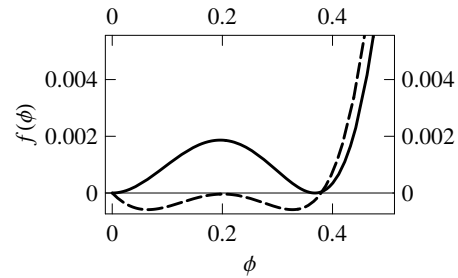}
\caption{\label{fig:2} The free energy at $T=0$ for $h=0$, $t=t_1$ (solid
curve) and $h=0.02$, $t=0.1878$ (dashed curve), respectively. In both cases,
$u=v=m_0=1$ and $T=0$.}
\end{figure}

We now turn to the properties of the equation of state for $h\neq 0$. Consider
first the $T=0$ plane. Suppose $t$ has been tuned to $t_1$, so that at $h=0$
the free energy has two degenerate minima, one at $\varphi=0$ and one at
$\varphi=m_1$. For a small $h>0$ this double-minimum structure persists, and
the two minima can still be made degenerate by increasing $t$, see Fig.\
\ref{fig:2}.
There thus still is a first-order transition.
However, with increasing $h$ the
two minima merge at a point where the first three derivatives of $f$ vanish:
$f'(\varphi_{\text{c}}) = f''(\varphi_{\text{c}}) = f'''(\varphi_{\text{c}}) =
0$ with $\varphi_{\text{c}} = m_0\,e^{-13/12}\,e^{-u/2v}$. This condition
determines a critical point $(t_{\text{c}},h_{\text{c}})$ in the $T=0$ plane
that marks the endpoint of a line of first-order transitions. One finds
\bse
\label{eqs:5}
\bea
t_{\text{c}} &=& 6\,v\,\varphi_{\text{c}}^2 = 6\,e^{-13/6}\,m_0^2\,v\,e^{-u/v},
\label{eq:5a}\\
h_{\text{c}} &=& \frac{16}{3}\,v\,\varphi_{\text{c}}^3 =
\frac{16}{3}\,e^{-13/4}\,m_0^3\,v\,e^{-3u/2v}.
\label{eq:5b}
\eea
\ese

\begin{figure}[t]
\includegraphics[width=5.0cm]{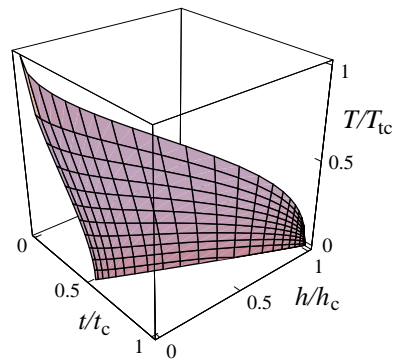}
\caption{\label{fig:3} The surface of first-order transitions in the space
spanned by $T$, $t$, and $h$. It is bounded by lines of first-order transitions
in the $T=0$ and $h=0$ planes, respectively, and by the line of second-order
transitions discussed after Eqs.\ (\ref{eqs:5}). A symmetric surface extends
into the region where $h<0$.}
\end{figure}
This discussion can be repeated for any fixed value of $T < T_{\text{tc}}$.
Accordingly, there is a line of critical points connecting the tricritical
point at $(T=T_{\text{tc}},t=0,h=0)$ and the quantum critical point at
$(T=0,t=t_{\text{c}},h=h_{\text{c}})$. A parametric representation for this
line is
\bse
\label{eqs:6}
\bea
t(s) &=& t_1\,(4s^2 + 5s + 6)\,(1-s)^2\,e^{1+g(s)},
\label{eq:6a}\\
T(s) &=& T_{\text{tc}}\,s^{1/2}\,e^{g(s)/2},
\label{eq:6b}\\
h(s) &=& h_{\text c}\,(s^2 + s + 1)\,(1-s)^{5/2}\,e^{13/4 + 3g(s)/2},
\label{eq:6c}
\eea
where
\be
g(s) = \frac{-1}{6}\,(4s^2 + 7s + 13)\,(1-s).
\ee
\ese
This line forms a boundary of a surface of first-order transitions that is
shown in Fig.\ \ref{fig:3}.

We see that the phase diagram obtained from Eq.\ (\ref{eq:3}) has the same
structure as the one observed experimentally, see Fig.\ \ref{fig:1}. As is the
case with Landau theory, the phase diagram in a space spanned by observables
will be a stretched and rotated version of the one in $T-t-h$ space, since the
parameters of the theory are complicated functions of pressure, temperature,
and magnetic field \cite{Chaikin_Lubensky_1995}. However, a quantitative
comparison with experiment can be made by expressing the zero-temperature
critical field strength $h_{\text{c}}$, Eq.\ (\ref{eq:5b}), in terms of
observable quantities, namely, the discontinuities of the magnetization and the
magnetic susceptibility, respectively, across the first-order transition at
$T=h=0$. The former is given by $m_1$, and differentiating the equation of
state shows the latter to be $\Delta\chi = 1/4vm_1^{\ 2}$. We find
\be
h_{\text{c}} = \frac{4}{3}\,e^{-7/4}\,m_1/\Delta\chi \approx
0.23\,m_1/\Delta\chi.
\label{eq:7}
\ee
A very rough estimate using the data from Ref.\ \onlinecite{Uhlarz_et_al_2004}
predicts a value of $h_{\text{c}}$ on the order of $0.1\ \text{T}$ for
ZrZn$_2$.

We now turn to the critical behavior at the various critical points in the
phase diagram. Equation (\ref{eq:3}) yields mean-field critical behavior for
all points on the lines of critical points at $T>T_{\text{tc}}$ and
$T_{\text{tc}}>T>0$. In particular, the order parameter critical exponents
$\beta$ and $\delta$ have their mean-field values $\beta=1/2$ and $\delta=3$,
respectively. At the tricritical point one finds mean-field tricritical
behavior \cite{Griffiths_1970, Griffiths_1973}, with $\beta=1/4$ and
$\delta=5$. This behavior gets modified if fluctuations are taken into account.
For $T>T_{\text tc}$ the upper critical dimension $d_{\text{c}}^+$, above which
mean-field critical behavior is exact, is $d_{\text{c}}^+=4$. For $d=3$ the
exact critical behavior is in the classical Heisenberg, XY, or Ising
universality class, depending on the nature of the ferromagnet. (ZrZn$_2$ is a
Heisenberg magnet; UGe$_2$ has a strong spin anisotropy and is thus Ising-like;
MnSi is a weak helimagnet, which leads to some complications which we ignore
here \cite{heli_footnote}.) For the wing-critical lines at $T<T_{\text{tc}}$,
$d_{\text{c}}^+=4$ as well. The exact critical behavior is always in the Ising
universality class, since the external magnetic field reduces the effective
dimension of the order parameter to one. At the tricritical point,
$d_{\text{c}}^+=3$, and the mean-field theory yields the exact critical
behavior except for logarithmic corrections to scaling
\cite{Wegner_Riedel_1973}.

For the quantum critical behavior at $T=0$ the mean-field theory yields the
usual mean-field values for the static exponents, e.g., $\beta=1/2$ and
$\delta=3$. For the temperature dependence of the order parameter at the
critical point one finds $\delta\varphi(t_{\text{c}},h_{\text{c}},T) \propto -
T^{2/3}$, where $\delta\varphi = \varphi - \varphi_{\text{c}}$. In the light of
Ref.\ \onlinecite{Belitz_et_al_2001a} one might expect the exact critical
behavior to differ strongly from these results. However, a detailed analysis
shows that Hertz theory \cite{Hertz_1976,Millis_1993} holds at this quantum
critical point. The reason is that the nonzero magnetic field and magnetization
suppress the soft-mode effects which invalidate Hertz theory, and ultimately
destroy the quantum critical point, at $h=0$. More generally, it was shown in
Ref.\ \onlinecite{Belitz_Kirkpatrick_Vojta_2002} that Hertz theory is valid if
the field conjugate to the order parameter does not change the soft-mode
structure of the system. In the present case, an expansion in powers of
$\delta\varphi$ about the quantum critical point shows that the quantity
$2\varphi_{\text{c}}\delta t - \delta h$, with $\delta t = t-t_{\text{c}}$ and
$\delta h = h - h_{\text{c}}$, plays the role of the conjugate field. Switching
on an external magnetic field from $h=0$ gives certain soft modes a mass, but
changing $h$ from $h_{\text{c}}\neq 0$ does not lead to further changes in the
soft-mode spectrum, and neither does changing the value of $t$. Mean-field
theory thus gives the exact static quantum critical behavior, in particular
\be
\beta = 1/2\quad,\quad \delta=3.
\label{eq:8}
\ee
However, the dynamic quantum critical behavior, i.e., the temperature
dependence at criticality, is modified from the mean-field result
\cite{Millis_1993,Sachdev_1997}, since the leading temperature dependence of
the parameter $t$ appears only at one-loop order. This fluctuation effect leads
to a temperature scale with a scale dimension $[T]_{\text{fluc}} = 9/(d+1)$.
For $d<5$ this dominates the Fermi-liquid temperature scale, which has
$[T]_{\text{FL}} = 3/2$ and is responsible for the temperature dependence of
the order parameter within mean-field theory. In $d=3$ we thus have the exact
result
\be
\delta\varphi(t_{\text c},h_{\text c},T) \propto -T^{4/9}.
\label{eq:9}
\ee
Notice that the static order parameter does not depend on the critical
temperature scale, which determines the dynamical critical exponent $z$ proper,
\be
z \equiv [T]_{\text c} = 3.
\label{eq:10}
\ee
For $d>2$, the critical scale dominates the fluctuation scale for all
observables that depend on it, e.g., the specific heat
\cite{Millis_1993,Sachdev_1997}. Notice that the above results are the {\em
exact} quantum critical behavior.

We finally discuss the relation between the theory presented above and a
competing mean-field theory with a very different microscopic underpinning.
Sandeman et al. \cite{Sandeman_Lonzarich_Schofield_2003} have proposed a Stoner
model where the equation of state is analytic in the order parameter, but band
structure effects, in particular a double-peak structure in the density of
states near the Fermi level, lead to signs of the coefficients consistent with
a first-order transition. These authors have shown that this provides an
explanation, not just for the first-order nature of the
paramagnet-to-ferromagnet transition, but also for a second, meta\-magnetic,
transition observed in the ferromagnetic phase of UGe$_2$, and they have argued
that it also leads to triplet superconductivity within the ferromagnetic state,
in agreement with observations on UGe$_2$ and URhGe. Band structure
calculations for these two materials have confirmed that a double-peak
structure near the Fermi level exists \cite{Shick_2004}. It is interesting to
compare various features and predictions of these two theories.

\noindent (1) The Stoner theory relies on detailed band-structure effects to
explain the first-order nature of the transition. The present theory, on the
other hand, is based on a universal many-body effect, namely, the existence of
soft particle-hole excitations, which are {\em always} present in metals.
It therefore predicts the first-order transition to be a {\em generic} feature
of low-$T_{\text{c}}$ itinerant ferromagnets, independent of the details of the
band structure.

\noindent (2) Within the Stoner theory one expects a temperature dependence of
the coefficient $u$ in Eq.\ (\ref{eq:9}) from Fermi liquid theory
\cite{Baym_Pethick_1991}, namely, $u = u_0 - u_1\,(T/T_0)^2$. Here $T_0$ is the
same microscopic temperature scale as in Eq.\ (\ref{eq:1}), and $u_0/u_1$ is on
the order of unity. One therefore expects $T_{\text{tc}}$ to be generically on
the order of $T_0$, and it is {\it a priori} not clear what suppresses
$T_{\text{tc}}$ to the observed values around 10K. The many-body theory, on the
other hand, provides a natural explanation for this effect: The coefficient $v$
in Eq.\ (\ref{eq:1}) reflects a mode-mode coupling effect, and therefore
$v/u\ll 1$ \cite{Kirkpatrick_Belitz_2003}. $T_{\text{tc}}$ is thus
exponentially small compared to $T_0$.

\noindent (3) Both theories yield magnetic-field dependences of the phase
diagram that are qualitatively the same, and quantitatively very close to one
another. For instance, the relation given by Eq.\ (\ref{eq:7}) is the same in
the Stoner theory, only the coefficient changes to $3\times
2^{7/2}/5^{5/2}\approx 0.61$. Notice that no magnetic field dependence of the
coefficients of either theory is necessary in order to produce the
characteristic ``wing structure'' of the phase diagram, the term $-h\phi$ in
the free energy suffices. In fact, the ``wing structure'' is a direct
consequence of the existence of the tricritical point \cite{Griffiths_1970} and
will be present in any theory that describes the latter.

We thank Christian Pfleiderer for stimulating discussions. This work was
supported by the NSF under grant Nos. DMR-01-32555 and DMR-01-32726 and by a
fellowship of the Deutsche Forschungsgemeinschaft (J.R.). Part of this work was
performed at the Aspen Center for Physics.


\begin{thebibliography}{24}
\expandafter\ifx\csname natexlab\endcsname\relax\def\natexlab#1{#1}\fi
\expandafter\ifx\csname bibnamefont\endcsname\relax
  \def\bibnamefont#1{#1}\fi
\expandafter\ifx\csname bibfnamefont\endcsname\relax
  \def\bibfnamefont#1{#1}\fi
\expandafter\ifx\csname citenamefont\endcsname\relax
  \def\citenamefont#1{#1}\fi
\expandafter\ifx\csname url\endcsname\relax
  \def\url#1{\texttt{#1}}\fi
\expandafter\ifx\csname urlprefix\endcsname\relax\def\urlprefix{URL }\fi
\providecommand{\bibinfo}[2]{#2} \providecommand{\eprint}[2][]{\url{#2}}

\bibitem[{\citenamefont{Jones and March}(1973)}]{Jones_March_1973}
\bibinfo{author}{\bibfnamefont{W.}~\bibnamefont{Jones}} \bibnamefont{and}
  \bibinfo{author}{\bibfnamefont{N.~H.} \bibnamefont{March}},
  \emph{\bibinfo{title}{Theoretical Solid State Physics}}
  (\bibinfo{publisher}{Wiley, London}, \bibinfo{year}{1973}).

\bibitem[{\citenamefont{Uhlarz et~al.}()\citenamefont{Uhlarz, Pfleiderer, and
  Hayden}}]{Uhlarz_et_al_2004}
\bibinfo{author}{\bibfnamefont{M.}~\bibnamefont{Uhlarz}},
  \bibinfo{author}{\bibfnamefont{C.}~\bibnamefont{Pfleiderer}},
  \bibnamefont{and} \bibinfo{author}{\bibfnamefont{S.~M.}
  \bibnamefont{Hayden}}, \eprint{cond-mat/0408424}.

\bibitem[{\citenamefont{Huxley et~al.}(2000)\citenamefont{Huxley, Sheikin, and
  Braithwaite}}]{Huxley_Sheikin_Braithwaite_2000}
\bibinfo{author}{\bibfnamefont{A.}~\bibnamefont{Huxley}},
  \bibinfo{author}{\bibfnamefont{I.}~\bibnamefont{Sheikin}}, \bibnamefont{and}
  \bibinfo{author}{\bibfnamefont{D.}~\bibnamefont{Braithwaite}},
  \bibinfo{journal}{Physica B} \textbf{\bibinfo{volume}{284-288}},
  \bibinfo{pages}{1277} (\bibinfo{year}{2000}).

\bibitem[{\citenamefont{Pfleiderer et~al.}(2001)\citenamefont{Pfleiderer,
  Julian, and Lonzarich}}]{Pfleiderer_Julian_Lonzarich_2001}
\bibinfo{author}{\bibfnamefont{C.}~\bibnamefont{Pfleiderer}},
  \bibinfo{author}{\bibfnamefont{S.~R.} \bibnamefont{Julian}},
  \bibnamefont{and} \bibinfo{author}{\bibfnamefont{G.~G.}
  \bibnamefont{Lonzarich}}, \bibinfo{journal}{Nature}
  \textbf{\bibinfo{volume}{414}}, \bibinfo{pages}{427} (\bibinfo{year}{2001}).

\bibitem[{hel()}]{heli_footnote}
\bibinfo{note}{Strictly speaking, MnSi is not a ferromagnet, but rather a weak
  helimagnet. This leads to complications in the detailed structure of the
  phase diagram \protect\cite{Thessieu_et_al_1997}, which we neglect here, as
  was done in Ref.\ \protect\onlinecite{Pfleiderer_Julian_Lonzarich_2001}.}

\bibitem[{\citenamefont{Hertz}(1976)}]{Hertz_1976}
\bibinfo{author}{\bibfnamefont{J.}~\bibnamefont{Hertz}},
  \bibinfo{journal}{Phys. Rev. B} \textbf{\bibinfo{volume}{14}},
  \bibinfo{pages}{1165} (\bibinfo{year}{1976}).

\bibitem[{\citenamefont{Millis}(1993)}]{Millis_1993}
\bibinfo{author}{\bibfnamefont{A.~J.} \bibnamefont{Millis}},
  \bibinfo{journal}{Phys. Rev. B} \textbf{\bibinfo{volume}{48}},
  \bibinfo{pages}{7183} (\bibinfo{year}{1993}).

\bibitem[{\citenamefont{Kimura et~al.}()\citenamefont{Kimura, Endo, Isshiki,
  Minagawa, Ochiai, Aoki, Terashima, Uji, Matsumoto, and
  Lonzarich}}]{Kimura_et_al_2004}
\bibinfo{author}{\bibfnamefont{N.}~\bibnamefont{Kimura}},
  \bibinfo{author}{\bibfnamefont{M.}~\bibnamefont{Endo}},
  \bibinfo{author}{\bibfnamefont{T.}~\bibnamefont{Isshiki}},
  \bibinfo{author}{\bibfnamefont{S.}~\bibnamefont{Minagawa}},
  \bibinfo{author}{\bibfnamefont{A.}~\bibnamefont{Ochiai}},
  \bibinfo{author}{\bibfnamefont{H.}~\bibnamefont{Aoki}},
  \bibinfo{author}{\bibfnamefont{T.}~\bibnamefont{Terashima}},
  \bibinfo{author}{\bibfnamefont{S.}~\bibnamefont{Uji}},
  \bibinfo{author}{\bibfnamefont{T.}~\bibnamefont{Matsumoto}},
  \bibnamefont{and} \bibinfo{author}{\bibfnamefont{G.~G.}
  \bibnamefont{Lonzarich}}, \bibinfo{journal}{Phys. Rev. Lett.}
  \textbf{\bibinfo{volume}{92}}, \bibinfo{pages}{197002} (????).

\bibitem[{\citenamefont{Kirkpatrick and
  Belitz}(2003)}]{Kirkpatrick_Belitz_2003}
\bibinfo{author}{\bibfnamefont{T.~R.} \bibnamefont{Kirkpatrick}}
  \bibnamefont{and} \bibinfo{author}{\bibfnamefont{D.}~\bibnamefont{Belitz}},
  \bibinfo{journal}{Phys. Rev. B} \textbf{\bibinfo{volume}{67}},
  \bibinfo{pages}{024419} (\bibinfo{year}{2003}).

\bibitem[{\citenamefont{Belitz et~al.}(2001)\citenamefont{Belitz, Kirkpatrick,
  Mercaldo, and Sessions}}]{Belitz_et_al_2001a}
\bibinfo{author}{\bibfnamefont{D.}~\bibnamefont{Belitz}},
  \bibinfo{author}{\bibfnamefont{T.~R.} \bibnamefont{Kirkpatrick}},
  \bibinfo{author}{\bibfnamefont{M.~T.} \bibnamefont{Mercaldo}},
  \bibnamefont{and} \bibinfo{author}{\bibfnamefont{S.~L.}
  \bibnamefont{Sessions}}, \bibinfo{journal}{Phys. Rev. B}
  \textbf{\bibinfo{volume}{63}}, \bibinfo{pages}{174427}
  (\bibinfo{year}{2001}).

\bibitem[{\citenamefont{Landau and Lifshitz}(1980)}]{Landau_Lifshitz_V_1980}
\bibinfo{author}{\bibfnamefont{L.~D.} \bibnamefont{Landau}} \bibnamefont{and}
  \bibinfo{author}{\bibfnamefont{E.~M.} \bibnamefont{Lifshitz}},
  \emph{\bibinfo{title}{Statistical Physics Part 1}}
  (\bibinfo{publisher}{Butterworth Heinemann, Oxford}, \bibinfo{year}{1980}).

\bibitem[{\citenamefont{Stoner}(1938)}]{Stoner_1938}
\bibinfo{author}{\bibfnamefont{E.~C.} \bibnamefont{Stoner}},
  \bibinfo{journal}{Proc. Roy. Soc. London A} \textbf{\bibinfo{volume}{165}},
  \bibinfo{pages}{372} (\bibinfo{year}{1938}).

\bibitem[{\citenamefont{Sandeman et~al.}(2003)\citenamefont{Sandeman,
  Lonzarich, and Schofield}}]{Sandeman_Lonzarich_Schofield_2003}
\bibinfo{author}{\bibfnamefont{K.}~\bibnamefont{Sandeman}},
  \bibinfo{author}{\bibfnamefont{G.}~\bibnamefont{Lonzarich}},
  \bibnamefont{and}
  \bibinfo{author}{\bibfnamefont{A.}~\bibnamefont{Schofield}},
  \bibinfo{journal}{Phys. Rev. Lett.} \textbf{\bibinfo{volume}{90}},
  \bibinfo{pages}{167005} (\bibinfo{year}{2003}).

\bibitem[{\citenamefont{Belitz et~al.}(1999)\citenamefont{Belitz, Kirkpatrick,
  and Vojta}}]{Belitz_Kirkpatrick_Vojta_1999}
\bibinfo{author}{\bibfnamefont{D.}~\bibnamefont{Belitz}},
  \bibinfo{author}{\bibfnamefont{T.~R.} \bibnamefont{Kirkpatrick}},
  \bibnamefont{and} \bibinfo{author}{\bibfnamefont{T.}~\bibnamefont{Vojta}},
  \bibinfo{journal}{Phys. Rev. Lett.} \textbf{\bibinfo{volume}{82}},
  \bibinfo{pages}{4707} (\bibinfo{year}{1999}).

\bibitem[{\citenamefont{Belitz et~al.}()\citenamefont{Belitz, Kirkpatrick, and
  Vojta}}]{Belitz_Kirkpatrick_Vojta_2004}
\bibinfo{author}{\bibfnamefont{D.}~\bibnamefont{Belitz}},
  \bibinfo{author}{\bibfnamefont{T.~R.} \bibnamefont{Kirkpatrick}},
  \bibnamefont{and} \bibinfo{author}{\bibfnamefont{T.}~\bibnamefont{Vojta}},
  \eprint{cond-mat/0403182}.

\bibitem[{\citenamefont{Chaikin and Lubensky}(1995)}]{Chaikin_Lubensky_1995}
\bibinfo{author}{\bibfnamefont{P.}~\bibnamefont{Chaikin}} \bibnamefont{and}
  \bibinfo{author}{\bibfnamefont{T.~C.} \bibnamefont{Lubensky}},
  \emph{\bibinfo{title}{Principles of Condensed Matter Physics}}
  (\bibinfo{publisher}{Cambridge University, Cambridge}, \bibinfo{year}{1995}).

\bibitem[{\citenamefont{Griffiths}(1970)}]{Griffiths_1970}
\bibinfo{author}{\bibfnamefont{R.~B.} \bibnamefont{Griffiths}},
  \bibinfo{journal}{Phys. Rev. Lett.} \textbf{\bibinfo{volume}{24}},
  \bibinfo{pages}{716} (\bibinfo{year}{1970}).

\bibitem[{\citenamefont{Griffiths}(1973)}]{Griffiths_1973}
\bibinfo{author}{\bibfnamefont{R.~B.} \bibnamefont{Griffiths}},
  \bibinfo{journal}{Phys. Rev. B} \textbf{\bibinfo{volume}{7}},
  \bibinfo{pages}{545} (\bibinfo{year}{1973}).

\bibitem[{\citenamefont{Wegner and Riedel}(1973)}]{Wegner_Riedel_1973}
\bibinfo{author}{\bibfnamefont{F.~J.} \bibnamefont{Wegner}} \bibnamefont{and}
  \bibinfo{author}{\bibfnamefont{E.~K.} \bibnamefont{Riedel}},
  \bibinfo{journal}{Phys. Rev. B} \textbf{\bibinfo{volume}{7}},
  \bibinfo{pages}{248} (\bibinfo{year}{1973}).

\bibitem[{\citenamefont{Belitz et~al.}(2002)\citenamefont{Belitz, Kirkpatrick,
  and Vojta}}]{Belitz_Kirkpatrick_Vojta_2002}
\bibinfo{author}{\bibfnamefont{D.}~\bibnamefont{Belitz}},
  \bibinfo{author}{\bibfnamefont{T.~R.} \bibnamefont{Kirkpatrick}},
  \bibnamefont{and} \bibinfo{author}{\bibfnamefont{T.}~\bibnamefont{Vojta}},
  \bibinfo{journal}{Phys. Rev. B} \textbf{\bibinfo{volume}{65}},
  \bibinfo{pages}{165112} (\bibinfo{year}{2002}).

\bibitem[{\citenamefont{Sachdev}(1997)}]{Sachdev_1997}
\bibinfo{author}{\bibfnamefont{S.}~\bibnamefont{Sachdev}},
  \bibinfo{journal}{Phys. Rev. B} \textbf{\bibinfo{volume}{55}},
  \bibinfo{pages}{142} (\bibinfo{year}{1997}).

\bibitem[{\citenamefont{Shick}(2004)}]{Shick_2004}
\bibinfo{author}{\bibfnamefont{A.~B.} \bibnamefont{Shick}}
  (\bibinfo{year}{2004}), \bibinfo{note}{talk given at the 20th General
  Conference of the Condensed Matter Division of the European Physical Society,
  Prague, Czech Republic, July 2004}.

\bibitem[{\citenamefont{Baym and Pethick}(1991)}]{Baym_Pethick_1991}
\bibinfo{author}{\bibfnamefont{G.}~\bibnamefont{Baym}} \bibnamefont{and}
  \bibinfo{author}{\bibfnamefont{C.}~\bibnamefont{Pethick}},
  \emph{\bibinfo{title}{Landau Fermi-Liquid Theory}}
  (\bibinfo{publisher}{Wiley, New York}, \bibinfo{year}{1991}).

\bibitem[{\citenamefont{Thessieu et~al.}(1997)\citenamefont{Thessieu,
  Pfleiderer, Stepanov, and Flouquet}}]{Thessieu_et_al_1997}
\bibinfo{author}{\bibfnamefont{C.}~\bibnamefont{Thessieu}},
  \bibinfo{author}{\bibfnamefont{C.}~\bibnamefont{Pfleiderer}},
  \bibinfo{author}{\bibfnamefont{A.~N.} \bibnamefont{Stepanov}},
  \bibnamefont{and} \bibinfo{author}{\bibfnamefont{J.}~\bibnamefont{Flouquet}},
  \bibinfo{journal}{J. Phys.: Cond. Matter} \textbf{\bibinfo{volume}{9}},
  \bibinfo{pages}{6677} (\bibinfo{year}{1997}).

\end{thebibliography}

\end{document}